\documentclass[prb,twocolumn,showpacs,a4paper]{revtex4}
\usepackage{graphics}
\usepackage{dcolumn}
\begin{document}
\title{Coupling to haloform molecules in intercalated C$_{60}$?}
\author{Erik Koch}
\email{E.Koch@fkf.mpg.de}
\author{Olle Gunnarsson}
\affiliation{Max-Planck Institut f\"ur Festk\"orperforschung,
             Heisenbergstra\ss e 1, 70569 Stuttgart, Germany}
\date{July 24, 2002}
\begin{abstract}
For field-effect-doped fullerenes it was reported that the superconducting
transition temperature $T_c$ is markedly larger for
C$_{60}\cdot$2CHX$_3$ (X=Cl, Br) crystals, than for pure C$_{60}$.
Initially this was explained by the expansion of the volume per 
C$_{60}$-molecule and the corresponding increase in the density of states
at the Fermi level in the intercalated crystals. On closer examination
it has, however, turned out to be unlikely that this is the mechanism 
behind the increase in $T_c$. An alternative explanation of the enhanced
transition temperatures assumes that the conduction electrons not only 
couple to the vibrational modes of the C$_{60}$-molecule, but also to
the modes of the intercalated molecules. 
We investigate the possibility of such a coupling.
We find that, assuming the ideal bulk structure of the intercalated crystal,
both a coupling due to hybridization of the molecular levels, and a coupling 
via dipole moments should be very small. This suggests that the presence
of the gate-oxide in the field-effect-devices strongly affects the structure
of the fullerene crystal at the interface. 
\end{abstract}
% 74.70.Wz   Fullerenes and related materials
\maketitle

In a series of papers Sch\"on and collaborators announced a number 
of amazing results: Doping pure C$_{60}$ using a field-effect device, they found
superconductivity up to temperatures of 11 K for electron\cite{elecdoped} 
and 52 K for hole-doping.\cite{holedoped}
Replacing pure C$_{60}$ by crystals intercalated with chloroform (CHCl$_3$)
and bromoform (CHBr$_3$), they reported transition temperatures of about
18 and 25 K for electron-doping, and 80 and 117 K for hole-doping, 
respectively.\cite{latticeexp}
Initially it was speculated that this increase in $T_c$ was due to the
expansion of the lattice upon intercalation of the CHX$_3$-molecules and
the correspondingly larger density of states (DOS) at the Fermi 
level\cite{latticeexp} --- a mechanism similar to that seen in the 
alkali doped fullerenes.\cite{rmp}

While the alkali doped fullerenes are bulk superconductors, the 
induced charge in the field-effect-doped materials is believed to be
essentially confined to a single C$_{60}$-monolayer.\cite{wehrli}
This monolayer is believed to be a [111]-layer for pure C$_{60}$ and a
[010]-layer for C$_{60}\cdot$2CHX$_3$.
Analyzing the low-temperature phase of the haloform intercalated fullerenes
shows, that the expansion of the unit cell volume induced by the 
intercalated molecules mainly results from an increase in the distance 
{\em between} these layers, while the density of states {\em in} these layers 
does not correlate with the observed $T_c$.\cite{jansen}
%Even if DOS increases as expected, cannot be explained in usual framework
%\cite{nonadiabatic}.

It was proposed that the transition temperature increases not
because of an enhanced density of states at the Fermi level, but because
of an additional coupling to the intercalated CHX$_3$ (X=Cl, Br) 
molecules.\cite{bill}
Here we ask if indeed there is such a coupling.
Possible mechanisms are (i) hybridization of the molecular levels
with the HOMO/LUMO of C$_{60}$ and (ii) coupling via the dipole moment
or the polarizability of the intercalated molecules.
We find that the overlap between the states of the intercalated 
molecules with the relevant orbitals of C$_{60}$ ($h_u$ for hole- and 
$t_{1u}$ for electron-doping) is very small. Moreover, group theory puts
rather strong constraints on this type of coupling.
The second coupling mechanism also does not seem to be viable, as 
electrostatic interactions should be strongly reduced by the efficient
screening found in the fullerenes.\cite{schluter,screening}
Since our results were obtained for the ideal bulk structure, the experimental
results\cite{elecdoped,holedoped,latticeexp} may imply that the presence 
of the gate oxide in the field-effect-device strongly affects the structure 
of the fullerene crystal at the interface.

\begin{widetext}
{\em Hybridization:}
To understand the electron-phonon coupling to the intercalated haloform 
molecules, we calculate the coupling constant\cite{Rainer} for a molecular
solid with more than one molecule per unit cell:
\begin{displaymath}
 \lambda={2\over N(0)}\sum_i\sum_{\nu_i,q}{1\over2M_{\nu_i}\Omega_{\nu_i,q}^2}
  \sum_{n,m;k}|\langle n,k|V_{\nu_i,q}|m,k+q\rangle|^2
   \delta(\varepsilon_{n,k})\delta(\varepsilon_{m,k+q}-\Omega_{\nu_i,q}) .
\end{displaymath}
Here $N(0)$ is the total density of states per spin at the Fermi level,
$i$ runs over the different molecules in the unit cell, $\nu_i$ labels the
vibrational modes of molecule $i$, and $q$ is the phonon wave vector.
Expanding the Bloch function $|n,k\rangle$ in molecular orbitals $\alpha_j$ 
on molecule $j$ at position $r_j$ in unit cell $R$
\begin{displaymath}
 |n,k\rangle = {1\over\sqrt{N}}\sum_{R,j}e^{ik(R+r_j)}
   \sum_{\alpha_j}c_{\alpha_j}^n(k)|\Phi_{R,j,\alpha_j}\rangle ,
\end{displaymath}
the electron-phonon matrix element is given by a sum over the matrix elements
$\langle\Phi_{R,j,\alpha_j}|V_{\nu_i,q}|\Phi_{R',j',\alpha'_{j'}}\rangle$.
For a molecular solid we can neglect the intermolecular electron-phonon
coupling\cite{Lannoo} and thus obtain
\begin{displaymath}
 \langle n,k|V_{\nu_i,q}|m,k+q\rangle =
 {1\over\sqrt{N}}\sum_{\alpha_i,\alpha_i'}
   \overline{c_{\alpha_i}^n(k)}\, c_{\alpha_i'}^m(k+q)\;
    V_{\alpha_i,\alpha_i'}(\nu_i) ,
\end{displaymath}
where $V_{\alpha_i,\alpha_i'}(\nu_i)$ is the electron-vibration matrix element
on molecule $i$. Writing the partial density of states as 
\begin{displaymath}
 n_{\alpha,\alpha'}(\varepsilon)={1\over N}\sum_{n,k}
  \overline{c_{\alpha}^n(k)}\, c_{\alpha'}^n(k)\;
  \delta(\varepsilon-\varepsilon_{n,k}) ,
\end{displaymath}
we finally obtain
\begin{displaymath}
 \lambda={2\over N(0)/N}\sum_i\sum_{\nu_i}{1\over2M_{\nu_i}\Omega_{\nu_i}^2}
  \sum_{\alpha_i,\alpha_i',\alpha_i'',\alpha_i'''}
   \overline{V_{\alpha_i,\alpha_i'}(\nu_i)}\,V_{\alpha_i'',\alpha_i'''}(\nu_i)\;
    n_{\alpha_i,\alpha_i''}(0)\, n_{\alpha_i',\alpha_i'''}(\Omega_{\nu_i}) .
\end{displaymath}
\end{widetext}
Thus a vibrational mode of a molecule only contributes to the electron-phonon
coupling, if (i) there is a molecular orbital that contributes significantly
to the density of states at the Fermi level and (ii) the electron-vibration
matrix element does not vanish. In the case of C$_{60}$, the relevant orbitals
are the $t_{1u}$ ($h_u$) for electron (hole) doping and the 
non-vanishing electron-vibration matrix elements matrix are found by 
reduction of the symmetric tensor product 
$t_{1u}\otimes_s t_{1u}=A_g\oplus H_g$ 
($h_u\otimes_s h_u=A_g\oplus G_g\oplus 2H_g$).

\begin{figure}[p]
 \resizebox{3in}{!}{\includegraphics{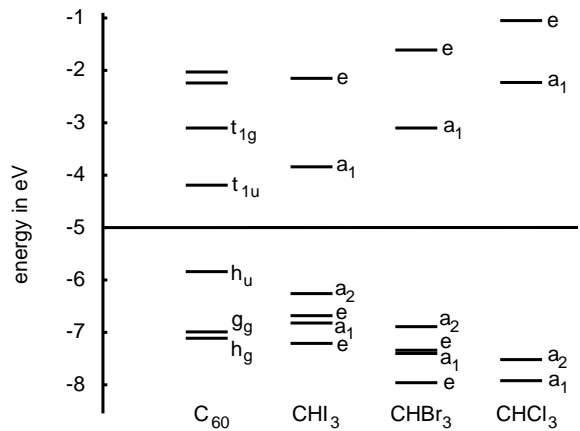}}
 \caption[]{\label{levels}
            Molecular levels of C$_{60}$ and CHX$_3$ (X=Cl, Br, I).
            Note that the HOMO and LUMO of the haloform molecules
            are singly degenerate (irreducible representations $a_1$ or $a_2$
            of the symmetry group $C_{3v}$).}
\end{figure}

To estimate the contribution of the molecular levels of the CHX$_3$-molecules
to the density of states at the Fermi level we have performed
all-electron density functional calculations using the Gaussian-orbital code 
NRLMOL,\cite{NRLMOL} employing the PBE functional.\cite{PBE} 
The basis set is 4s3p1d for H, 5s4p3d for C, 6s5p3d for
Cl, and 7s6p4d for Br.  The position of the energy levels of CHX$_3$
(X=Cl, Br, I) compared to those of C$_{60}$ are shown in figure \ref{levels}.
While the levels of the chloroform molecule are fairly distant from the 
HOMO/LUMO levels of C$_{60}$, the levels of bromoform and, in particular,
iodoform move much closer. This implies that the contribution of the haloform
orbitals to the density of states at the Fermi level should increase when
replacing chlorine by bromine, suggesting an explanation of the increase in
transition temperature.

To estimate the actual contribution of the CHX$_3$ levels to the density of 
states at the Fermi level, we perform calculations both for 
an isolated C$_{60}$-molecule and for a system consisting of a 
C$_{60}$-molecule and the twelve closest neighboring CHX$_3$-molecules at the 
experimentally determined positions.\cite{jansen} We then calculate
$\sum_{m,n}\langle\Psi_m|\Phi_n\rangle\langle\Phi_n|\Psi_m\rangle$,
where the $\Psi_m$ are the $t_{1u}$ ($h_u$) derived orbitals for the
C$_{60}$-molecule with the neighboring CHX$_3$-molecules and the
$\Phi_n$ are the $t_{1u}$ ($h_u$) orbitals of the isolated C$_{60}$-molecule.
If there were no hybridization between C$_{60}$ and CHX$_3$, the overlap
would be equal to the number of C$_{60}$-derived orbitals considered
(3 for $t_{1u}$ and 5 for $h_u$). The deviation from this number is a measure
of the hybridization between C$_{60}$ and CHX$_3$.
As shown in table \ref{overlaps}, we find that the deviation is very small 
for both CHCl$_3$ and CHBr$_3$, i.e., there is essentially no hybridization of 
the haloform-levels with the HOMO/LUMO of the C$_{60}$-molecule: Less than
3\% for C$_{60}\cdot$2CHBr$_3$ and less than 1.5 \% for C$_{60}\cdot$2CHCl$_3$.
Only the $g_g$ and $h_g$ levels that are well below the Fermi level 
show appreciable hybridization, as could be expected from the energetic 
proximity of these levels and the occupied levels of CHCl$_3$ and CHBr$_3$.
Therefore, the contribution of the CHX$_3$-levels to the electron-phonon 
coupling should be very small.

The situation changes, of course, in the field-doped layer. There will be
an additional electrostatic potential, which can lead to a shift in the
relative positions of the electronic levels of the C$_{60}$ and the 
intercalated haloform molecules. One might then speculate that for a 
certain external field one can line up the HOMO or LUMO of the different
molecules, thereby maximizing the mixing and consequently a possible
coupling to the haloform modes. For that field one would then expect to
find the maximum transition temperature.
Since the energetic positions of the molecular levels of CHCl$_3$ and CHBr$_3$
are quite different (cf.~figure \ref{levels}), the fields required for 
bringing say the HOMO of CHCl$_3$ in line with that of C$_{60}$ is 
substantially larger than that required for CHBr$_3$. Hence one would expect 
that the transition temperature for C$_{60}\cdot$2CHCl$_3$ peaks at a 
gate-voltage significantly different from that for C$_{60}\cdot$2CHBr$_3$. 
Since the gate-voltage also corresponds to the induced charge carrier density,
from the above scenario one would expect that the transition temperatures
for the different crystals would show a maximum at different doping levels
-- contrary to the experimental finding reported in reference
\onlinecite{latticeexp}.

In addition group theory puts further constraints on this coupling via
selection rules for the electron-vibration matrix elements. As seen 
from figure \ref{levels}, the HOMO/LUMO levels of the haloform molecules 
are singly degenerate of symmetry $a_2$ and $a_1$, respectively. Decomposing 
the (symmetric) tensor product of these irreducible representations, we find 
that they can only couple to the molecular vibrations of symmetry $A_1$, not to
the two-fold degenerate $E$ modes (cf.~tables \ref{modes} and \ref{coupling}). 
Only the two-fold degenerate molecular levels of symmetry $e$, which are even 
further away from the Fermi level than the HOMO/LUMO, can couple to all the 
modes. Hence even if there is some contribution of the HOMO/LUMO levels of the 
CHX$_3$ molecules to the density of states at the Fermi level, coupling to 
the majority of molecular modes would be forbidden by symmetry.

{\em Electrostatic coupling:}
Due to the dipole moment of the haloform molecules 
(3.4\,10$^{-30}$ Cm for CHCl$_3$ and 3.3\,10$^{-30}$ Cm for CHBr$_3$)
one might speculate that there is coupling to the intercalated molecules
due to electrostatic interactions. An analogous scenario was put forward
for the case of the alkali-doped fullerene A$_3$C$_{60}$, where it was
suggested early on that the superconductivity is mediated by coupling to the
vibrations of the alkali ions.\cite{zhang} Experimentally, however, no
isotope effect was found for the alkalis.\cite{isotope} 
This could be explained as a consequence of the efficient screening in
the alkali-doped fullerenes, which leads to a strong reduction in the
coupling to the alkali-modes.\cite{schluter,screening}
A similar mechanism should be at work in the field-doped fullerenes,
reducing the coupling to the dipole moments of the haloform molecules.
In addition, one would expect, that even the unscreened coupling to 
dipoles (haloform molecules) should be weaker than the coupling to
monopoles (alkali ions). Moreover, the dipole moments of CHCl$_3$ and
CHBr$_3$ are very similar, while, because of the lattice expansion,
the bromoform molecules are more distant from the C$_{60}$ than the
chloroform molecules. Based on a coupling to the dipole moment, one
would therefore expect that $T_c$ in C$_{60}\cdot$2CHBr$_3$ should be
lower than in C$_{60}\cdot$2CHCl$_3$ -- contrary to the experimental
finding. A coupling via the dipole moments thus seems very unlikely.

{\em Conclusions:}
In most of the arguments we have given above, we have assumed that
the structure and symmetry of the bulk crystal is also relevant
in the region under the gate oxide. Our results show that with this
assumption it seems hard to understand the observed increase in
transition temperature of the haloform intercalated fullerenes.
This then suggests that the crystal structure under the gate oxide must be
markedly different from the bulk structure. Possible effects are
a different orientation of the molecules at the interface, a reduction
of the symmetry, some bonding to the oxide, some additional screening
due to the presence of the oxide, or a reorientation of the haloform molecules
in the strong electric field used for in field-doping.
It therefore seems that a correct picture that allows to understand the
experimental results reported in reference \onlinecite{latticeexp} must 
involve more than just assuming that a monolayer of the ideal crystal is 
doped with charge carriers.

{\em Acknowledgments}
We would like to thank A.~Burkhard for helpful discussions and 
M.R.~Pederson and J.~Kortus for their support in using the NRLMOL code.

\begin{table}[p]
 \caption[]{Overlap 
            $\sum_{m,n}\langle\Psi_m|\Phi_n\rangle\langle\Phi_n|\Psi_m\rangle$
            between the molecular orbitals $\Psi_m$ of C$_{60}$ with the twelve
            closest neighboring CHX$_3$ (X=Cl, Br) molecules and the 
            molecular orbitals $\Phi_n$ of the isolated C$_{60}$ molecule.
            The first column gives 
            $\sum_{m,n}\langle\Phi_m|\Phi_n\rangle\langle\Phi_n|\Phi_m\rangle$,
            the degeneracy of the levels.}
 \label{overlaps}
 \begin{ruledtabular}
 \begin{tabular}{l|cllc}
  MO & C$_{60}$ & CHBr$_3$ & CHCl$_3$\\    % only the 2 CHX3 that are in basis
  \hline
  $t_{1g}$ & 3  & 0.97     & 2.93\\        % 2.816   &  2.972\\
  $t_{1u}$ & 3  & 2.91     & 2.96 & LUMO\\ % 2.993   &  2.994 & LUMO\\
  $h_u$    & 5  & 4.88     & 4.95 & HOMO\\ % 4.974   &  4.991 & HOMO\\
  $g_g$    & 4  & 2.30     & 3.20\\        % 2.658   &  3.784\\
  $h_g$    & 5  & 1.61     & 4.04          % 2.906   &  4.744
 \end{tabular}
 \end{ruledtabular}
\end{table}

\begin{table}[p]
 \caption[]{Vibrational modes of CHX$_3$ molecules and their symmetry.\cite{LB}}
 \label{modes}
 \begin{ruledtabular}
 \begin{tabular}{l|cccccc}
  meV      & $E$   & $A_1$ & $A_1$ & $E$ & $E$ & $A_1$\\
  \hline
% CHCl$_3$ & 261 & 363 & 672 & 760 & 1217 & 3030\\
% CHBr$_3$ & 154 & 222 & 539 & 656 & 1142 & 3023
  CHCl$_3$ & 32.5 & 45.4 & 82.8 & 94.4 & 150.8 & 374.4\\
  CHBr$_3$ & 19.1 & 27.5 & 66.8 & 81.3 & 141.6 & 374.9
 \end{tabular}
 \end{ruledtabular}
\end{table}

\begin{table}[p]
 \caption[]{Electron-phonon coupling for the HOMO and LUMO of the CHX$_3$ 
            molecule: reduction of the symmetric tensor product into irreducible
            representations. Coupling between different molecular orbitals:
            reduction of the tensor product. The electrons can only couple to
            the two-fold degenerate vibrational modes ($E$) when a two-fold
            degenerate molecular level ($e$) is involved.} % (Hamermesh, pp.132)
 \label{coupling}
 \begin{ruledtabular}
 \begin{tabular}{c@{\hspace{2ex}}c|rrr|c}
        & $C_{3v}$ & $E$ & $2C_3$ & $3c_v$\\
   \hline
   LUMO & $a_1 \otimes_s a_1$ & 1 & 1 & 1 & $A_1$\\
   HOMO & $a_2 \otimes_s a_2$ & 1 & 1 & 1 & $A_1$\\
        & $e   \otimes_s e  $ & 3 & 0 & 1 & $A_1 \oplus E$\\[0.8ex]
        & $a_1 \otimes   a_1$ & 1 & 1 & 1 & $A_1$\\
        & $a_1 \otimes   a_2$ & 1 & 1 &-1 & $A_2$\\
        & $a_1 \otimes   e  $ & 2 &-1 & 0 & $E  $\\
        & $a_2 \otimes   a_2$ & 1 & 1 & 1 & $A_1$\\
        & $a_2 \otimes   e  $ & 2 &-1 & 0 & $E  $\\
        & $e   \otimes   e  $ & 4 & 1 & 0 & $A_1 \oplus A_2 \oplus E$
 \end{tabular}
 \end{ruledtabular}
\end{table}

\begin{thebibliography}{}
\bibitem{elecdoped}
 J.H.~Sch\"on, Ch.~Kloc, R.C.~Haddon, and B.~Batlogg,
 Science {\bf 288}, 656 (2000).
\bibitem{holedoped}
 J.H.~Sch\"on, Ch.~Kloc, and B.~Batlogg, Nature {\bf 408}, 549 (2000).
\bibitem{latticeexp}
 J.H.~Sch\"on, Ch.~Kloc, and B.~Batlogg, Science {\bf 293}, 2432 (2001).
\bibitem{rmp}
 O.~Gunnarsson, Rev.~Mod.~Phys.~{\bf 69}, 575 (1997).
\bibitem{wehrli}
 S.~Wehrli, D.~Poilblanc, and T.M.~Rice, 
 Eur.~Phys.~J.~B {\bf 23}, 345 (2001). 
\bibitem{jansen}
 R.E.~Dinnebier, O.~Gunnarsson, H.~Brumm, E.~Koch, P.W.~Stephens, A.~Huq, 
 and M.~Jansen, Science {\bf 296}, 109 (2002).
%\bibitem{nonadiabatic}
% P.~Paci, E.~Cappelluti, C.~Grimaldi, L.~Pietronero, and S.~Str\"assler,
% {\tt cond-mat/0201334}.
\bibitem{bill}
 A.~Bill and V.Z.~Kresin, Eur.~Phys.~J.~B {\bf 26}, 3 (2002);
 A.~Bill, R.~Windiks, B.~Delley, and V.Z.~Kresin,
 Int.~J.~Mod.~Phys.~{\bf 16}, 1533 (2002).
\bibitem{schluter}
 M.~Schl\"uter, M.~Lannoo, M.~Needels, G.A.~Baraff, and D.~Tom\'{a}nek,
 J.~Phys.~Chem.~Solids {\bf 53}, 1473 (1992).
\bibitem{screening}
 E.~Koch, O.~Gunnarsson, and R.M.~Martin,
 Phys.~Rev.~Lett.~{\bf 83}, 620 (1999).
\bibitem{Rainer}
 D.~Rainer, Prog.~Low Temp.~Phys.~{\bf 10}, 371 (1986).
\bibitem{Lannoo}
 M.~Lannoo, G.A.~Baraff, M.~Schl\"uter, and D.~Tomanek,
 Phys.~Rev.~B {\bf 44}, 12106 (1991).
\bibitem{NRLMOL}
 M.R.~Pederson and K.A.~Jackson, Phys.~Rev.~B.~{\bf 41}, 7453 (1990);
 K.~Jackson and M.R.~Pederson, Phys.~Rev.~B.~{\bf 42}, 3276 (1990);
 M.R.~Pederson and A.A.~Quong, Phys.~Rev.~B {\bf 46}, 13584 (1992);
 A.A.~Quong, M.R.~Pederson, and J.L.~Feldman,
  Solid State Commun.~{\bf 87}, 535 (1993);
 D.~Porezag and M.R.~Pederson, Phys.~Rev.~A.~{\bf 60}, 2840 (1999).
\bibitem{PBE}
 J.P.~Perdew, K.~Burke and M.~Ernzerhof, Phys.~Rev.~Lett.~{\bf 77}, 3865 (1996).
\bibitem{LB}
 Landoldt-B\"ornstein, Vol.~I/2, table 14145 IX b), Springer, Heidelberg, 1951.
\bibitem{zhang}
 F.C.~Zhang, M.~Ogata, and T.M.~Rice, Phys.~Rev.~Lett.~{\bf 67}, 3452 (1991).
\bibitem{isotope}
 B.~Burk, V.H.~Crespi, A.~Zettl, and M.L.~Cohen,
 Phys.~Rev.~Lett.~{\bf 72}, 3706 (1994).
\end{thebibliography}
\end{document}